# AN EFFICIENT GROUP AUTHENTICATION FOR GROUP COMMUNICATIONS

Lein Harn[1] and Changlu Lin[2]

[1]Department of Computer Science Electrical Engineering, University of Missouri-Kansas City, MO 64110, USA

`harnl@umkc.edu`

[2]Key Laboratory of Network Security and Cryptology, Fujian Normal University, Fujian, 35007, P. R. China

`cllin@fjnu.edu.cn`

## ABSTRACT

*Group communication implies a many-to-many communication and it goes beyond both one-to-one communication (i.e., unicast) and one-to-many communication (i.e., multicast). Unlike most user authentication protocols that authenticate a single user each time, we propose a new type of authentication, called* **group authentication**, *that authenticates all users in a group at once. The group authentication protocol is specially designed to support group communications. There is a group manager who is responsible to manage the group communication. During registration, each user of a group obtains an unique token from the group manager. Users present their tokens to determine whether they all belong to the same group or not. The group authentication protocol allows users to reuse their tokens without compromising the security of tokens. In addition, the group authentication can protect the identity of each user.*

## 1. INTRODUCTION

User authentication is one of the most important security services in computer and communication application. Knowledge based authentication (e.g., password) [16,9] and key based authentication (e.g., public/private key) [7,12] are the two most popular approaches. Knowledge based authentication has some security flaws. Most users like to use simple and short passwords. However, Internet hackers can easily crack simple passwords. Public-key based authentication needs a certificate authority (CA) to provide the authenticity of public keys. In addition, public-key computations involve large integers. Computational time is one of the main concerns for public-key based authentication.

All user authentication protocols [10,6] are one-to-one type of authentication where the prover interacts with the verifier to prove the identity of the prover. For example, the RSA digital signature [13] is used to authenticate the signer of the signature. In this approach, the verifier sends a random challenge to the prover. Then, the prover digitally signs the random challenge and returns the digital signature of the challenge to the verifier. After successfully verifying the digital signature, the verifier is convinced that the prover is the one with the identity of the public key used to verify the digital signature. In wireless communications, when a mobile subscriber wants to establish a connection with the base station, the subscriber and the base station interact to establish mutual authentication. Mutual authentication can prevent an illegitimate subscriber from using the service and prevent the fake base station from harming the subscriber.

Network applications are no longer just one-to-one communication; but involve multiple users $(>2)$. Group communication [14,2] implies a many-to-many communication and it goes beyond both one-to-one communication (i.e., unicast) and one-to-many communication (i.e., multicast). In this paper, we propose a new type of authentication, called ***group authentication***, which authenticates all users in a group at once. The group authentication protocol is specially designed to support group communications. The group authentication is defined to involve multiple users and users want to convince each other that they all belong to the same group without revealing their identities. In the group authentication, each user acts as both the prover and the verifier. Group authentication is extremely important in an ad hoc network because this network is temporarily established by multiple users and these users want to use this network to exchange secret information.

Devising protocols to provide group authentication in ad hoc networks is extremely challenging due to highly dynamic and unpredictable topological changes. As a result, there are two popular models to provide group authentication services in an ad hoc network. The first model involves a centralized authentication server (AS) [11,3] and the second model has no AS [5,4]. In the first model, AS manages the access rights of the network. For example, Bhakti et al. [3] proposed to adopt Extensible Authentication Protocol (EAP) in the IEEE 802.1x standard for wireless ad hoc network. This approach requires to set up the AS and have mobile users to access to the AS service. In fact, in some situations, the second model is the only way to provide group authentication. For example, in an ad-hoc network communication, there has no AS service available to mobile users. In the second model, each user needs to take in charge of authenticating other users. In a straightforward approach, if there are $n$ users in the group, each user can use the one-to-one authentication protocol for $n-1$ times to authenticate other users. Computational time is one of the major concerns in this approach.

In this paper, we introduce a special type of group authentication which provides an efficient way to authenticate multiple users belonging to the same group without revealing identity of each user. Our proposed protocol is no longer a one-to-one type of authentication. It is a many-to-many type of authentication. Unlike most user authentication protocols that authenticate a single user each time, our proposed protocol authenticates all users of a group at once. In our proposal, each user needs to register with a group manager (GM) to become a group user. Like the trusted dealer in Shamir's $(t,n)$ secret sharing scheme [15], the GM needs to select a secret polynomial and compute token for each user. Based on these tokens, our protocol can establish group authentication for all users at once. The group authentication protocol allows users to reuse their tokens without compromising the security of tokens. Our proposed protocol supports existing wireless communication network including wireless ad hoc network.

**The rest of this paper is organized as follows.** In next section, we include some preliminaries. In Section 3, we introduce the model of our proposed group authentication. In Section 4, we present basic one-time group authentication protocol; in Section 5, we present group authentication protocol without revealing tokens. We conclude in Section 6.

## 2. PRELIMINARIES

### 2.1. Review of Shamir's secret sharing scheme [15]

In Shamir's $(t,n)$ secret sharing scheme based on the polynomial, there are $n$ shareholders and a mutually trusted dealer. The scheme consists of two algorithms:

a) **Share generation algorithm:** the dealer first picks a random polynomial of degree $t-1$, $f_i(x) = a_{t-1}x^{t-1} + \cdots + a_1 x + a_0 \pmod{p}$, such that the secret $s$ satisfies $f(0) = a_0 = s$ and all coefficients, $a_0, a_1, \ldots, a_{t-1} \in Z_P$, $p$ is a prime with $p > s$. The dealer computes shares, $f(x_i)$, for $i = 1, 2, \ldots, n$, and distributes each share $f(x_i)$ to shareholder $U_i$ secretly.

b) **Secret reconstruction algorithm:** it takes any $t$ or more than $t$ shares, for example, $j$ shares (i.e., $t \leq j \leq n$), $(x_1, f(x_1)), (x_2, f(x_2)), \ldots, (x_j, f(x_j))$, as inputs, and outputs the secret $s$ using Lagrange interpolating formula as

$$s = \sum_{i=1}^{j} f(x_i) \prod_{r=1, r \neq i}^{j} \frac{-x_r}{x_i - x_r} \pmod{p}.$$

We note that the above algorithms satisfy the basic requirements of the secret sharing scheme, that are, (1) with the knowledge of any $t$ or more than $t$ shares, shareholders can reconstruct the secret $s$; and (2) with the knowledge of any $t-1$ or fewer than $t-1$ shares, shareholders cannot obtain the secret $s$. Shamir's secret sharing scheme is unconditionally secure since the scheme satisfies these two requirements without making any computational assumption. For more information on this scheme, please refer to the original paper [15].

### 2.2. Harn and Lin's definition on strong $t$-consistency [8]

Benaloh [1] presented a notion of $t$-consistency to determine whether a set of shares is generated from a polynomial of degree $t-1$ at most. Recently, Harn and Lin [8] proposed a new definition of strong $t$-consistency which is the extension of Benaloh's definition.

**Definition 1 (Strong $t$-consistency [8]).** A set of $n$ shares (i.e., $t < n$) is said to be strong $t$-consistent if (a) any subset of $t$ or more than $t$ shares can reconstruct the secret, and (b) any subset of fewer than $t$ shares cannot reconstruct the secret.

It is obvious that if shares in Shamir's secret sharing scheme are generated by a polynomial with degree $t-1$ exactly, then shares satisfy the security requirements of a $(t,n)$ secret sharing scheme and these shares are also strong $t$-consistent.

Checking strong $t$-consistency of $n$ shares can be executed very efficiently by using Lagrange interpolating formula. In fact, to check whether $n$ shares are strong $t$-consistent or not, it only needs to check whether the interpolation of $n$ shares yields a polynomial with degree $t-1$ exactly. If this condition is satisfied, we can conclude that all shares are strong $t$-consistent. However, if there are some illegitimate shares, the degree of the interpolating polynomial of these $n$ shares is more than $t-1$ with very high probability. In other words, these $n$ shares are most likely to be not strong $t$-consistent. The property of strong $t$-consistency will be used in Section 5 of our protocol to check strong $t$-consistency of $n$ shares without revealing tokens.

## 3. MODEL

### 3.1. Entities

a) **Group Manager (GM):** A group manager is responsible to register users to form a group. The responsibility of GM is to issue a secret token to each user during registration. Later, authentication is based on the secret tokens. Since tokens are used in authentication, identities of users are protected. In order to prevent malicious users to reveal their tokens to attackers, each token is a unique integer. The secret tokens are shares of the polynomial generated by the GM.

b) **Group Users:** Join a group and become a group user, each user needs to register with the GM. After being successfully registered, each user receives a secret token from the GM. Each user with a unique token can prevent malicious users to give their tokens to impersonators.

c) **Attackers:** We consider two types of attackers, the *inside attackers* and the *outside attackers*. The inside attackers are users who are legitimate users and own legitimate tokens from the GM. We consider that the insider attackers may collude to forge tokens for non-users. The outside attackers are impersonators who do not own any tokens and try to impersonate users to fail the authentication protocol. We also assume that the GM does not collude with any user. If the GM colludes with any user by revealing the secret of the GM to the user, the colluded user can do harm to the group. In addition, we assume all users act honestly in the authentication. If any use acts dishonestly by revealing a invalid value, the authentication is failed.

### 3.2. Authentication outcomes

There are only two possible outcomes of a group authentication; that are, either "yes" or "no". If the outcome is "yes", it means that all users belong to the same group; otherwise, there are impersonators.

## 4. BASIC ONE-TIME GROUP AUTHENTICATION PROTOCOL

In the following discussion, we assume that there are $n$ users, $M_1, M_2, \ldots, M_n$, registered at the GM to form a group.

### 4.1. System set up

During registration, GM constructs a random $(t-1)$-th (i.e., $t < n$) degree polynomial $f(x)$ with $f(0) = s$, and computes secret tokens of users as $y_i = f(x_i)$, for $i = 1, 2, \ldots, n$, where $x_i$ is the public information associated with user $M_i$. GM sends each token $y_i$ to user $M_i$ secretly. GM makes $H(s)$ publicly known, where $H$ is a one-way function.

*Remark 1.* The threshold $t$ is an important security parameter that affects the security of group authentication protocols. Using a $(t, n)$ secret sharing scheme to issue tokens in the registration can prevent up to $t - 1$ inside attackers, who are legitimate users, colluded together to forge tokens.

### 4.2. Basic one-time group authentication protocol

From now on, we assume that there are $j$ users with their tokens $f(x_1), f(x_2), \ldots, f(x_j)$ where $t < j \leq n$, who want to execute the group authentication protocol.

The basic idea of this protocol is that each user releases the token obtained from the GM during registration. If all released tokens are valid, the interpolation of the released tokens can reconstruct the secret $s$. The published one-way hash of the secret is used to compare with the one-way hash of the reconstructed secret.

**Theorem 1.** *Protocol 1 can detect any number of illegitimate users.*

*Proof.* If there is illegitimate user who does not own a valid token on the polynomial $f(x)$, the reconstructed secret will be different from the secret $s$. Thus, Protocol 1 can detect any number of illegitimate users. □

---

**Protocol 1:** One-time group authentication protocol

- **Step 1.** Each user $M_i$ reveals his token $f(x_i)$, to all other users simultaneously.

- **Step 2.** After knowing all tokens, $f(x_i)$, for $i = 1, 2, \ldots, j$, following Lagrange interpolating formula, each user computes $s' = \sum_{i=1}^{j} f(x_i) \prod_{r=1, r \neq i}^{j} \frac{-x_r}{x_i - x_r}$ $(\mod p)$. If $H(s') = H(s)$, all users have been authenticated successfully; otherwise, there are illegitimate users.

---

*Remark 2.* This is a one-time authentication protocol since the secret and tokens are revealed to all users in this protocol. The authentication is no longer a one-to-one authentication and it is a many-to-many authentication. The proposed protocol is very efficient to authenticate multiple users belonging to the same group without revealing identity of each user.

## 5. GROUP AUTHENTICATION PROTOCOL WITHOUT REVEALING TOKENS

In Protocol 1, since tokens are revealed to all users, each token can only be used for one-time authentication. In addition, the secret $s$ is also exposed to users in Protocol 1. In the following discussion, we propose a way to protect tokens. In addition, the secret does not need to be recovered in each authentication. Our authentication is based on the property of strong $t$-consistency in Section 2.2.

### 5.1. Group authentication protocol without revealing tokens

In the following protocol, it can be achieved authentication without revealing tokens and the secret. The basic idea of our approach uses the property of strong $t$-consistency. Let each user select a random polynomial with $(t-1)$-th degree and generate shares for other users. Then, each user releases the additive sum of his own token obtained from the GM during the registration and sum of shares of polynomials generated by users. Due to the property of secret

---

**Protocol 2:** Group authentication protocol without revealing tokens

- **Step 1.** Each user $M_i$ selects a random polynomial, $f_i(x)$, with $(t-1)$-th degree. For the polynomial $f_i(x)$, user $M_i$ computes shares as $f_i(x_r)$, for $r = 1, 2, \ldots, j, r \neq i$, for other users. User $M_i$ sends each share, $f_i(x_r)$ to user $M_r$ secretly.

- **Step 2.** After receiving $f_r(x_i)$ for $r = 1, 2, \ldots, j$, each user uses his token $f(x_i)$ to compute $y_i'' = f(x_i) + \sum_{r=1}^{j} f_r(x_i) \pmod{p}$. Each user releases his value $y_{i''}$.

- **Step 3.** After knowing $y_i''$, for $i = 1, 2, \ldots, j$, each user checks whether they are strong $t$-consistent. If they are not strong $t$-consistent, there are illegitimate users; else, all users have been successfully authenticated belonging to the same group.

---

sharing homomorphism in Section 2.2, the released sums are shares of the secret polynomial $f(x)$ of tokens and sum of polynomials generated by users. If all users act honestly and own valid tokens, the released sums should be strong $t$-consistent; otherwise, the released sums are not strong $t$-consistent. Since users do not need to reconstruct the secret in the protocol and the tokens have not been revealed directly, the dealer does not need to publish the one-way of the secret $s$ during system set up and the tokens can be reused.

**Theorem 2.** *Protocol 2 can detect any number of illegitimate users.*

*Proof.* Due to the property of secret sharing homomorphism, each released value, $y_i''$ in Step 2 is the share of additive sum of polynomials, $f(x) + \sum_{r=1}^{j} f_r(x) \pmod{p}$, with $(t-1)$-th degree. Thus, in Step 3, all released values, $y_i''$, for $i = 1, 2, \ldots, j$, are strong $t$-consistent. If there is any illegitimate user who does not own a valid token, $f(x_i)$, the released values, $y_i''$, for $i = 1, 2, \ldots, j$, are not strong $t$-consistent with very high probability. □

*Remark 3.* In Step 2, the token $f(x_i)$ cannot be computed from the revealed value $y_i'' = f(x_i) + \sum_{r=1}^{j} f_r(x_i) \pmod{p}$. Therefore, the tokens are protected unconditionally and can be reused for multiple authentications.

## 5.2. Computational complexity

The most time-consuming operation for each user is to check the strong $t$-consistency of released values $y_i''$ for $i = 1, 2, \ldots, j$, in Step 3 of Protocol 2. Following our discussion presented in Section 2.2, checking strong $t$-consistency needs to compute the interpolating polynomial of values $y_i''$. The polynomial interpolation becomes the main computational task in our proposed protocol. However, the modulus $p$ in our polynomial interpolation is much smaller than the

14

modulus in most public-key cryptosystems, such as RSA cryptosystem [13]. In addition, not like conventional user authentication protocol that authenticates one user at a time, this proposed authentication protocol authenticates all users at once. Thus, the proposed authentication protocol is very efficient in comparing with all existing authentication protocols.

## 6. CONCLUSIONS

We propose a special type of group authentication which is specially designed for group communications such as the ad hoc wireless communication network. The proposed group authentication protocol is no longer a one-to-one type of user authentication and it is a many-to-many type of authentication that authenticates multiple users at once. We first propose an basic one-time group authentication protocol and then propose a general group authentication protocol without revealing tokens. Our proposed group authentication is very efficient since the computation is based on the computation of linear polynomial.

## ACKNOWLEDGEMENTS

This research is supported by the National Natural Science Foundations of China under Grant No. 61103247 and the Natural Science Foundation of Fujian Province under Grant No. 2011J05147.

**Authors**

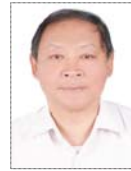

**Lein Harn** received the B.S. degree in electrical engineering from the National Taiwan University in 1977, the M.S. degree in electrical engineering from the State University of New York-Stony Brook in 1980, and the Ph.D. degree in electrical engineering from the University of Minnesota in 1984. In 1984, he joined the Department of Electrical and Computer Engineering, University of Missouri- Columbia as an assistant professor, and in 1986, he moved to Computer Science and Telecommunication Program (CSTP), University of Missouri, Kansas City (UMKC). While at UMKC, he went on development leave to work in Racal Data Group, Florida for a year. His research interests include cryptography, network security, and wireless communication security. He has published a number of papers on digital signature design and applications and wireless and network security. He has written two books on security. He is currently investigating new ways of using secret sharing in various applications.

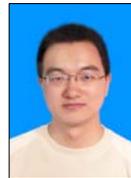

**Changlu Lin** received the BS degree and MS degree in mathematics from the Fujian Normal University, P.R. China, in 2002 and in 2005, respectively, and received the Ph.D degree in information security from the state key laboratory of information security, Graduate University of Chinese Academy of Sciences, P.R. China, in 2010. He works currently for the School of Mathematics and Computer Science, and the Key Laboratory of Network Security and Cryptology, Fujian Normal University. He is interested in cryptography and network security, and has conducted research in diverse areas, including secret sharing, public key cryptography and their applications.